\documentclass[prd,nofootinbib,twocolumn,showpacs]{revtex4}
\usepackage{graphicx}
\usepackage{bm}

\def\be{\begin{equation}}
\def\ee{\end{equation}}
\def\ba{\begin{eqnarray}}
\def\ea{\end{eqnarray}}

\begin{document}

\title{\large \bf Bounds on Cosmic Strings from WMAP and SDSS}
\author{Mark Wyman$^{1,2}$}\email{wyman@astro.cornell.edu}
\author{Levon Pogosian$^3$}
\author{Ira Wasserman$^{1,2}$ \vspace{0.2cm}}
\affiliation{$^1$ Laboratory for Elementary Particle Physics,
Cornell University,
Ithaca, NY 14853, USA \\
$^2$ Center for Radiophysics and Space Research, Cornell
University, Ithaca, NY 14853, USA \\
$^3$ Institute of Cosmology, Department of Physics and Astronomy,
Tufts University, Medford, MA 02155, USA}
\date{today}


\begin{abstract}

We find the constraints from WMAP and SDSS data on the fraction of
cosmological fluctuations sourced by local cosmic strings using a
Markov Chain Monte Carlo (MCMC) analysis. In addition to varying the usual 6
cosmological parameters and the string tension ($\mu$), we also varied
the amount of small-scale structure on the strings. Our results
indicate that cosmic strings can account for up to $7$ ($14$)\% of
the total power of the microwave anisotropy at $68$ ($95$)\% confidence level. 
The corresponding bound on the string mass per unit length, within our string model, 
is $G\mu < 1.8 (2.7) \times 10^{-7}$ at $68$ ($95$)\% c.l., where
this constraint has been altered from what appears below following the correction
of errors in our cosmic string code outlined in a recent erratum, astro-ph/0604141 \cite{newerratum}.
We also calculate the B-type polarization spectra sourced by cosmic strings and
discuss the prospects of their detection. 
\end{abstract}


\
\maketitle

NOTE: The results of the analysis presented in this paper are somewhat
modified in light of the discovery of errors in the computer code we
use in this paper. These errors, and their effects on the results,
are discussed in a new erratum to this paper, astro-ph/0604141 \cite{newerratum}.
The text of this paper has not been altered. Please read it together
with our erratum to learn what is still valid and what has changed.

\section{Introduction}

Cosmic strings were proposed as potential seeds for structure formation
in the early $1980$'s \cite{Zeldovich80,Vilenkin81a} and extensively studied 
in this context until the late $1990$'s (see \cite{VSbook} for a review). 
The interest was fueled, in part, by the fact that the CMB temperature anisotropy
discovered by COBE \cite{cobe} was of the same order of magnitude 
($10^{-5}$) as the density fluctuations that would be produced by cosmic strings formed 
around GUT epoch. However, by 1997 it was already apparent that strings could not explain
the distribution of large-scale structure \cite{ABR97}. Finally, when
Boomerang \cite{boomerang} and Maxima \cite{maxima} revealed the existence of
acoustic peaks in the CMB angular spectrum, strings were excluded as a viable
alternative to inflation as a model for seeding the formation of structure. The shape and location
of the peaks, currently measured to a high precision by WMAP \cite{wmap_bennett},
strongly indicate that structures grew out of initial fluctuations with adiabatic
initial conditions and with a scale-invariant spectrum, just as
prescribed by the inflationary paradigm \cite{guth}. In contrast, strings
predict
nearly featureless CMB spectra\footnote{
This is true for all ``conventional'' cosmic strings networks.
Technically, it is possible 
to construct active sources of fluctuations leading to CMB temperature spectra 
in agreement with observations, if, for example, one allows the time- and
space-correlations of strings to violate the bounds imposed by causality 
\cite{turok96,AM00}.} and, when normalized to COBE or WMAP, do not
produce enough clustering on large scales.

While strings could not have seeded all of the structure in the Universe,
they could have created a subdominant yet
non-negligible fraction of the primordial cosmological fluctuations 
\cite{jeannerot,kofman,tkachev,contaldi,battye,bouchet,mairi}. This idea
has recently received renewed attention with the realization that cosmic strings 
are produced in a wide class of string theory models of the inflationary epoch \cite{jst,costring,jst2,DV03,mcgill,hassan,Polchinski03, chen}.
In these models, inflation can arise during the collisions of branes that
coalesce to form, ultimately, the brane on which we live
\cite{dvali-tye,burgess,rabadan,collection}. 
Brane inflation predicts adiabatic temperature
and dark matter fluctuations capable of reproducing all currently
available observations. In addition, the collision at the conclusion of brane
inflation can produce a network of local cosmic
strings \cite{jst,costring}, whose effects on cosmological
observables range from negligible to substantial, depending on the
specific scenario \cite{jst2,DV03}.
It has also been shown that strings could form at an observationally acceptable 
level at the end of the D-term inflation in SUSY GUT models \cite{mairi}. As the
precision of cosmological observations increases, one might hope to
distinguish among the numerous presently-viable models of
inflation by studying and constraining the properties of the cosmic strings they predict.

The properties of the strings
produced at the end of brane inflation are similar to those of
local strings. 
They intercommute and break off loops 
like networks of usual strings, implying that they
will settle into a scaling solution. 
However, because of their higher-dimensional nature, they are expected to 
intercommute at a reduced rate and thus approach a scaling solution
at a higher string density. There are also other, qualitatively
new aspects of cosmic superstrings,
the most important of which is that string theory models
generically predict the existence of at least two fundamentally different kinds of cosmic strings.
These are the so-called fundamental, or $F$-strings, and another kind of string formed by
wrapping of all but one of the dimensions of a
 higher-dimensional $D$-brane around compact dimensions, or $D$-strings. The 
 relative tensions of these string types are determined by the superstring
 coupling, $g_s$. Because these
 different string types interact through binding, rather than intercommutation,
they can form higher-tension $(p,q)$ states composed of $p$ $F$-strings and
$q$ $D$-strings \cite{Polchinski03,Jackson04}. A first model of these networks has been developed recently,
and shows that these binding reactions allow network scaling \cite{us05}.
Because of the complexity introduced by this kind of new physics,
the amount of small-scale structure on strings 
 -- the string ``wiggliness" -- can vary depending on
 the particular brane inflation scenario. To make our constraints
applicable to a wide range of models, we have included the string
wiggliness as one of the parameters in the MCMC simulation. The effect of a
higher string density can be roughly modeled by appropriately adjusting the 
normalization of the string-generated spectra. This then translates into 
a simple adjustment of the bound on $G\mu$.

The aim of this paper is to constrain the properties of cosmic strings by
using the power spectrum data from the WMAP and SDSS experiments. There are
other ways to constrain cosmic strings -- some of them promising to
produce much tighter bounds than those that will ever be possible with
power spectrum data. We give a brief review of other methods in the
summary section, Sec.~\ref{summary}.

The rest of this paper is organized as follows. In
Sec.~\ref{methods} we give a detailed account of the model and the
methods used. We show the results in Sec.~\ref{results} and
conclude with a discussion in Sec.~\ref{summary}.

\section{The Model and Analysis}
\label{methods}

The fluctuations resulting from brane inflation are expected to be an
incoherent superposition of contributions from adiabatic perturbations
initiated by curvature fluctuations and active perturbations
induced by the decaying cosmic string network. The resulting
CMB angular spectra can be written as a sum of the adiabatic
and string contributions:
\be
C_l=C_l^{\rm ad}+C_l^{\rm cs}.
\label{clsuperpos}
\ee
Analogous expressions hold for matter density spectra. 
We restrict our study to a flat FRW universe and
vary the following cosmological parameters: the Hubble constant $h$,
the matter density $\Omega_M h^2$, the baryon density $\Omega_b h^2$, 
and the reionization optical depth $\tau$. In addition, we vary the galaxy bias
factor $F_b$, the amplitude $A_s$ and the spectral index $n_s$ of the primordial 
scalar power spectrum, as well as the string mass per unit length, $G\mu$,
and the string wiggliness parameter $\alpha_r$ (to be defined in Sec.~\ref{stringmodel}).

\begin{figure}
\centering
\includegraphics[width=80mm]{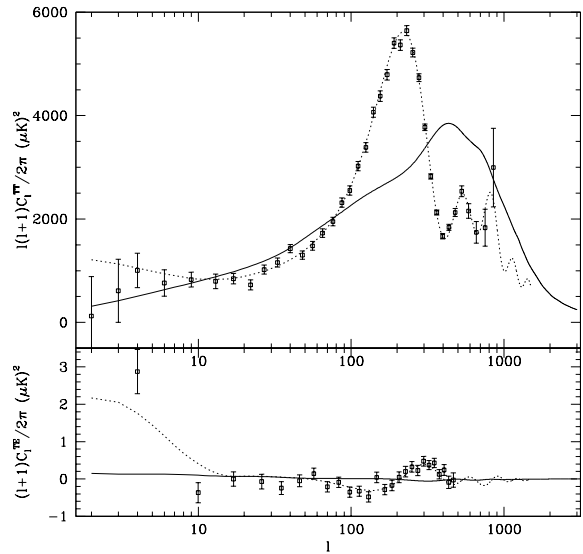}
\caption{\label{stringcl} The CMB TT and TE spectra (solid lines) sourced by cosmic strings
with wiggliness parameter $\alpha_r=1.9$, as well as the adiabatic spectra for the same cosmological 
parameters (dashed lines) and WMAP's first year data. The string spectra are normalized so 
that the {\it total} TT power is the same for the two lines, which corresponds to $B=1$.
}
\end{figure}

\begin{figure}
\centering
\includegraphics[width=80mm]{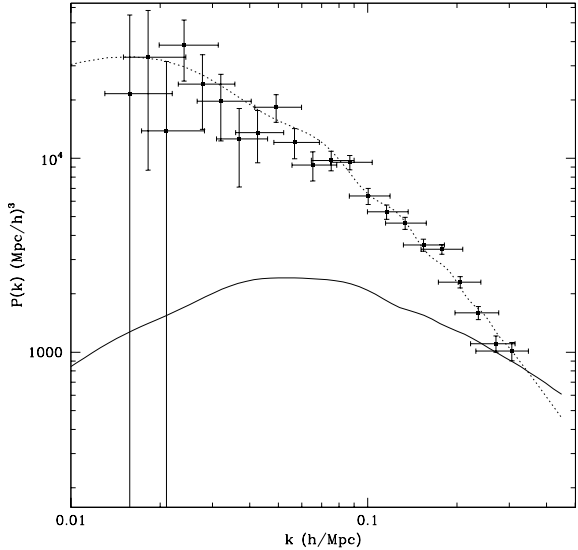}
\caption{\label{stringpk}The string-generated matter power spectrum (solid line) for the same parameters as 
in Fig.~\ref{stringcl}, i.e., for $B=1$. The dashed line represents the linear spectrum from adiabatic perturbations at $z=0$.}
\end{figure}

We used a suite of different codes to produce and analyze the spectra. 
The model we employed for the cosmic string-generated perturbations is 
described in the subsection below. The string CMB and matter spectra were calculated using 
a modification \cite{PV99,levon_web} of CMBFAST \cite{cmbfast} (see
Figs.~\ref{stringcl} and \ref{stringpk} for representative string induced spectra). 
We first evaluated and stored the string spectra on a grid of parameters. During
our calculations, the  spectra were obtained by interpolation
on the grid. The adiabatic matter spectra were also stored on a 
grid after having been evaluated using a publicly available version of CMBFAST.
For the adiabatic CMB spectra, we used the package CMBWarp \cite{cmbwarp}.

To compare the theoretical linear matter power spectrum $P^{\rm L}(k)$ generated by 
our code with the SDSS results, we first applied the halo-fitting 
procedure of Smith et al. \cite{halofit} to obtain the non-linear spectra,
$P^{\rm NL}(k)$.  
This procedure is only valid for the adiabatic contribution
to the $P(k)$ spectrum, so using it for the cosmic string
contribution introduces some inaccuracy into our model.
However, as we can see in Fig. \ref{stringpk}, the string power
spectrum is considerably weaker than
the adiabatic power spectrum on all but the smallest length
scales even in models where strings account for all the large-scale microwave anisotropy.
Thus, since only a small portion of the relevant values of $k$ are
affected by the halo-fitting procedure,  and since the effect is
small on the scales we consider, we expect any inaccuracy
introduced by our approximate treatment of nonlinearity to be
unimportant.
Next, we assumed that the
galaxy bias factor and the correction due to peculiar velocities on the
scales of our interest ($k < 0.2$ h/Mpc) are approximately redshift- and scale-independent
and can be combined into a single constant factor, $F_b$, multiplying the
matter spectrum:
\be
P(k)=F_b P^{\rm NL}(k) .
\label{pkfinal}
\ee
We then fit $P(k)$ to the SDSS power spectrum data (for $k < 0.2$ h/Mpc)
with $F_b$ as a free parameter. 

The likelihood of the spectra for a given set of parameters was obtained using 
the publicly-available likelihood codes produced by the WMAP \cite{wmaplike} 
and SDSS \cite{maxwebsite} teams. 

\subsection{The Cosmic String Model}
\label{stringmodel}

Unlike the adiabatic inflationary perturbations, which are set as
initial conditions in the equations of motion of linear
perturbations, cosmic strings act as a continual source of
fluctuations as the universe evolves.
The density and temperature
fluctuations created in the immediate locality of a cosmic string
are highly non-linear, e.g. the density contrast $\delta
\rho/\rho$ is significantly larger than unity in the wake formed
behind a moving string. However, the effect of strings on
cosmological scales is that of a small perturbation to the
evolution of the average cosmic energy-momentum tensor. Hence, for
the purpose of computing the CMB and LSS spectra, the metric and
density perturbations caused by strings are described by the
system of linearized Einstein-Boltzmann equations with strings
acting as active sources.

Evaluation of the CMB and LSS spectra sourced by strings
requires knowing their energy-momentum tensor (or its unequal time
correlation functions \cite{pen_ps}) for the entire dynamical
range of the calculation, which is approximately four orders of
magnitude in scale factor. Realistic simulations of cosmic
string networks have so far been limited either in their dynamical
range \cite{AWSA98} or their resolution \cite{LS04}. Hence, until full
scale simulations become available, one is forced to resort to
approximate methods to model the string sources.

Numerical simulations \cite{BenBouch,AllenShellard90,AlTur89} show
that during the radiation and matter dominated eras the string
network evolves according to a scaling solution, which on
sufficiently large scales can be described by two length scales.
The first scale, $\xi(t)$, is the coherence length of strings,
i.~e. the distance beyond which directions along the string are
uncorrelated. The second scale, $L(t)$, is the average
inter-string separation. Scaling implies that both length scales
grow in proportion to the horizon. Simulations indicate that $\xi
(t) \sim t$, while $L(t) = \gamma t$, with $\gamma \approx 0.8$ in
the matter era \cite{BenBouch,AllenShellard90}. The
one-scale model \cite{Vilenkin81b,Kibble85}, in which the two length
scales are assumed to be equal, has been quite successful in
describing the general properties of cosmic string networks
inferred from numerical simulations. These simulations
assume that cosmic strings reconnect on every intersection.
It is of interest to us, however, to also consider the case when
the reconnection probability is less than one. If strings can move
and interact in extra dimensions then, while appearing to
intersect in our three dimensions, they may actually miss each
other. Hence, the effective intercommutation rate of these strings
will generally be less than unity.  As a consequence, one would
expect more strings per horizon in these theories \cite{jst,DV03}.
Because of the straightening of wiggles on sub-horizon scales due
to the expansion of the universe, one would still expect $\xi(t)
\sim t$. However, the string density would increase, therefore
reducing the inter-string distance. Hence, smaller
inter-commutation probabilities imply smaller $\gamma$.

In addition, numerical simulations show that long strings possess
a great deal of small-scale structure in the form of kinks
and wiggles on scales much smaller than the horizon. To an observer
who cannot resolve this structure, the string will appear to be
smooth, but with a larger effective mass per unit length ${\tilde
\mu}$ and a smaller effective tension ${\tilde T}$. An unperturbed
string (with $\mu=T$) exerts no gravitational force on nearby
particles. In contrast, a wiggly string with ${\tilde \mu} >
{\tilde T}$ attracts particles like a massive rod. The effective
equation of state of a wiggly strings is ${\tilde \mu} {\tilde T}
= \mu^2$ \cite{Carter,AV90}. Depending on the brane inflation model, 
the presence of extra dimensions could mean that even more small scale 
structure would be present on the strings \cite{DV03}.

To calculate the sources of perturbations we use an updated
version of the cosmic string model first introduced by Albrecht et
al.\cite{ABR97} and further developed in refs. \cite{PV99,gangui}, where
the wiggly nature of strings was taken into account. In this
model, the string network is represented by a collection of
uncorrelated straight string segments produced at some early epoch,
moving with uncorrelated, random velocities. At every
subsequent epoch, a certain fraction of the number of segments
decays in a way that maintains network scaling. The length of each
segment at any time is taken to be equal to the correlation length
of the network. This length and the root-mean-square (r.m.s.) velocity of segments
are computed from the velocity-dependent one-scale model of
Martins and Shellard \cite{MS96}. The positions of segments are
drawn from a uniform distribution in space, and their orientations
are chosen from a uniform distribution on a two-sphere.
This model is a rather crude approximation of a realistic string
network. However, there are good reasons 
to believe that its predictions for the CMB and LSS spectra are
close to what one would obtain by using full-scale simulations.  
Its parameters have been calibrated to produce source correlation functions
in agreement with those in \cite{vincent}, where
comparison to a full simulation was possible. 
Also, the {\it shape} of the spectra obtained using this
model are in good agreement with results of other groups
\cite{magetal96,carlo99,dani}, who used different methods that are also approximate. 
Finally, on the cosmological scales probed by the CMB measurements, the
fine details of the string evolution do not play a major role. It is
the large-scale properties -- such as the scaling distance, the
equation of state (wiggliness), the r.m.s. velocity, and how all
these characteristics evolve through the radiation-matter equality -- that 
determine the shape of the string-induced spectra.
All of these effects are accounted for in our model and can, in principle, be
adjusted to match any specific cosmic string model.
We choose to work with this model because one can easily calculate the
sources for different cosmological parameters and because it
allows us to include the effect of the wiggliness \cite{PV99},
which could be one of the distinguishing features of strings
produced in theories with extra dimensions \cite{DV03}. The other
main effect of the presence of extra dimensions, the increased
string density, can be approximately factored in by multiplying the 
spectra by $N_s \sim \gamma^{-2}$.

For technical details of the model, the reader is referred to
\cite{PV99} and references therein. The wiggly
nature of strings is accounted for by modifying the string
energy-momentum tensor so that it corresponds to the wiggly string
equation of state:
\be \tilde \mu = \alpha \mu \ , \
\tilde T = \alpha^{-1} \mu  \ ,\ee
where $\alpha$ is a parameter describing the wiggliness,
$\tilde \mu$ and $\tilde T$
are the mass per unit length and the string tension of the wiggly
string, and $\mu$ is the tension (or, equivalently, the mass per unit length) of 
the smooth string. In
addition to modifying the equation of state, the presence of small-scale 
structure slows strings down on large scales. We account for this by dividing 
the root mean squared string velocity by the parameter $\alpha$.
The wiggliness of the strings remains approximately constant during the
radiation and matter eras, but changes its value during the transition between the two.
We take the radiation era value, $\alpha_r$, to be a free parameter that we vary,
and set the matter era value to be $\alpha_m=(1+\alpha_r)/2$, with a smooth
interpolation between the two values (as prescribed in \cite{PV99}).
For conventional strings, this roughly agrees
with results of numerical simulations \cite{BenBouch,AllenShellard90} which show a decrease
from $\alpha_r \sim 1.8-1.9$ in the radiation era to $\alpha_m \sim 1.4-1.6$ in the matter era.

The way the wiggliness modifies the shape of the spectra can be understood
qualitatively, once one examines the main physical processes that contribute
to CMB anisotropy on various scales. The spectrum can be approximately divided into
two main components. 
One is a roughly flat (scale-invariant) 
component that arises 
from the combined Kaiser-Stebbins effect \cite{KS_effect} 
of many strings in a scaling string network; 
this component is produced {\it after} last scattering. 
The second component is the bump peaked at $\ell \sim 450$, which is primarily
determined by the state of density and velocity perturbations {\it at the time}
of last scattering. CMB photons experience Doppler shifts when they (last-)scatter off 
the velocity flows created by the string wakes that exist at this epoch \cite{Doppler_effect}. 
In addition, all the
density fluctuations that existed at the time of last scattering are imprinted on the CMB
via the Sachs-Wolfe effect \cite{SW_effect}; these density fluctuations
are caused by the CDM wakes created by strings during the time between 
radiation-matter equality and last scattering. 
If we change the string substructure by 
introducing string wiggliness, there is a two-fold effect on the flat part
of the spectrum that arises from the Kaiser-Stebbins effect.
The first effect is simple: by nature, heavier strings have larger deficit angles, 
and thus produce larger CMB discontinuities. However, the size of the discontinuities 
is also proportional to string velocity and, since wiggly strings move more slowly
than straight ones,
the net increase in the size of the discontinuities is mitigated. The wakes behind 
a wiggly string are, nonetheless, always more prominent. This is because
of wiggly strings having a non-zero Newtonian potential
that attracts matter particles in the string's vicinity. The combined effect
of wiggliness on the flat component of the spectrum is an overall
increase in its amplitude, but a relatively weak one.
The peak, on the other hand, is directly enhanced, since the
wakes at the last scattering are more prominent for wiggly strings.
The net result is an overall increase in the amplitude of the spectra
coupled with an enhancement of the peak at $\ell\sim 450$.
The characteristic scale of the fluctuations 
at decoupling is set by the size of the typical wake, or, in turn,
by the coherence length $\xi(t)$ of the string network at that time.  
Independent of the string intercommutation probability,
the coherence length $\xi(t)$ is 
always expected to be roughly equal to $t$. This is 
because of the straightening of wiggles on subhorizon scales caused by the
expansion of the universe.
Therefore, the qualitative features of the spectrum are quite independent of
the details of the model. Hence, any observational constraint that we obtain based only
on the {\it shape} of the spectrum (such as the constraint on the parameter 
$f$ defined below) can be viewed as less model-dependent than our other results.

It is well known that properties and possible observational
signatures of global and local strings can be very
different. Global strings predict almost no power on small angular
scales for CMB temperature anisotropy \cite{pen_ps}, while local
strings -- as we have argued above -- produce a quite significant broad peak at $l\sim 450$ 
in a spatially flat universe \cite{magetal96,ABR97,PV99,carlo99,dani,LP00}; this can
 be seen in Fig.~\ref{stringcl}.
Global strings also induce a significantly larger vector
component of metric perturbations \cite{pen_vector}. Consequently, their prediction
for the strength of the $B$-type polarization \cite{pen_pol} is
generally higher than that of local strings. We will only concern
ourselves with local strings. For the most recent constraints on
global strings the reader is referred to \cite{bevis04} 
and \cite{fraisse}.


Rather than working directly with the string parameter $G\mu$, we introduce a
parameter $B$, defined as
\be
B \equiv \left({\mu \over \mu_0} \right)^2 \ ,
\label{defineB}
\ee
where $\mu_0$ is the tension that
one obtains by setting the total power in
string induced CMB anisotropy to be equal to the total power observed by WMAP. That is, we set
\be
I^{cs} \equiv \sum_l \frac{(2l+1)}{4\pi} C_l^{\rm cs}(\mu_0,\alpha^{(0)}_r,{\vec p}_0) = I^{\rm WMAP} \ ,
\label{totalpower}
\ee 
where we take $\alpha^{(0)}_r=1.9$ and ${\vec p}_0$ is a fixed set of the remaining 
cosmological parameters that have taken to correspond to the the best fit $\Lambda$CDM model
of Tegmark et al \cite{tegmark}. The value of $G\mu_0$ that we obtain with this prescription
is $2 \times 10^{-6}$. Speaking very loosely,
$B$ can be said to measure the fraction of the anisotropy due to strings.
Note, however, that this meaning is modified if, for instance, the strings have a different
amount of small-scale structure ($\alpha_r \ne 1.9$)
or if they have reduced intercommutation probabilities. $B$ is only really useful as an
intermediate parameter. Our main results
are the constraints on $G\mu$ that we obtain from $B$ and 
the fraction $f$ of the total CMB anisotropy due to strings (Eq.~(\ref{definef})).

\subsection{Markov Chain Monte Carlo}

Because of the large size of our parameter space (nine parameters in total), we have used 
the Markov Chain Monte Carlo (MCMC) method for exploring the likelihood surface and for 
generating marginalized posterior distribution functions for the model parameters. 
We employed the MCMC algorithm described in the Appendix of \cite{tegmark}. 

We ran eight separate chains 
initialized 
at randomly generated initial positions within our prior range. The priors,
given in Table \ref{priors}, were 
chosen with the expectation of the string
contribution being subdominant and the best fit parameters being close
to their WMAP best fit values \cite{wmap_spergel}\footnote{The restriction on 
wiggliness, $\alpha_r<4$, was to save computing time.}. Since we expected 
values of $B$ near zero to be preferred, we also allowed $B$ to range slightly
below zero so as not to restrict artificially the ability of the chain's random steps 
to explore near $B=0$; values
below zero were discarded when the data were analyzed. One advantage to our use of 
multiple chains rather than a single, long chain was that we were able to verify directly 
that there was 
adequate mixing in each chain, since each successfully forgot its starting location
and located the same maximum likelihood region of the parameter space.


\begin{table}
\begin{center}
\begin{tabular}{|c|} \hline
$0 \leq$  $B$ \\
$1 \leq$  $\alpha_r$  $\leq 4$ \\
$0 \leq$  $A_s$ \\
$0.92 \leq$  $n_s$  $\leq 1.07$ \\
$0.019 \leq$  $\Omega_B h^2$  $\leq 0.028$ \\
$0.1 \leq $  $\Omega_M h^2$  $\leq 0.2$\\
$0.5 \leq$  $h$  $\leq 0.8$ \\
$0 \leq$  $\tau$  $\leq 0.23$ \\
$0 \leq  $ $F_b$ \\
\hline
\end{tabular}
\end{center}
\caption{\label{priors} Prior constraints on the parameters}
\end{table}

\section{Results}
\label{results}
\subsection{The fraction in strings and $G\mu$}
To test our MCMC code we first ran a chain with the string contribution set
to zero ($B=0$). The results are shown in Table~\ref{tab:results}. Our results
are consistent with those found in a similar analysis of the same data by 
members of the SDSS team \cite{tegmark}.

\begin{figure}
\centering
\includegraphics[width=80mm]{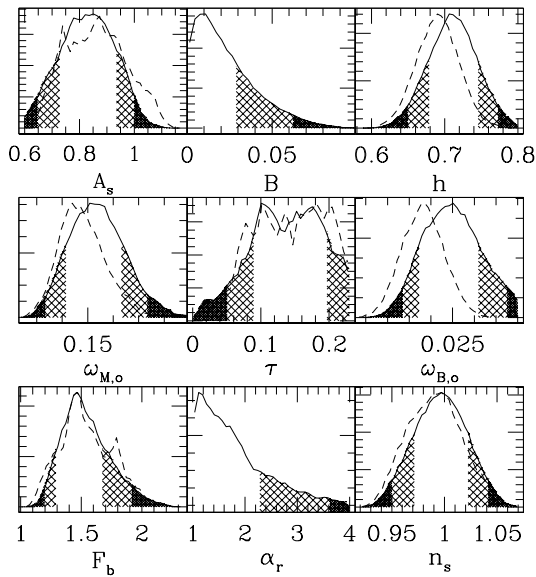}
\caption{\label{allmarg} The one-dimensional projected PDFs for the 9 parameters 
varied by our Markov Chain Monte Carlo code; note that $\omega_B = \Omega_B h^2$, 
$\omega_M = \Omega_M h^2$. The solid line represents the PDFs for models where cosmic 
strings are included; the dashed line represents the PDFs for models with $B=0$, 
i.e. without cosmic strings. Each curve has been rescaled such that 
its area is unity. For each PDF the lightly shaded regions are excluded at the 68\% 
confidence region; the dark regions are excluded at the 95\% confidence region.}
\end{figure}

\begin{table}
\begin{center}
\begin{tabular}{|ccc|}
\hline
Parameter & $B=0$ & $B>0$ \\
\hline
$f$  & --- & $<0.068\,(68\%)$, $<0.14\,(95\%)$ \\
$B$ & --- & $<0.029\, (68\%)$, $<0.062\,(95\%)$ \\
$\alpha_r$ & --- & $<2.3\,(68\%)$, $<3.6\, (95\%)$ \\
$A_s$ & $0.87^{+0.08}_{-0.16}$ &  $0.85^{+0.09}_{-0.13}$ \\
$n_s $ & $1.0^{+0.02}_{-0.04}$ & $1.0\pm0.026$ \\
$\Omega_B h^2$ & $0.024\pm 0.001$ &$0.025^{+0.0012}_{-0.0016}$ \\
$\Omega_M h^2$ & $0.15 \pm 0.01$ & $0.15^{+0.013}_{-0.01}$\\
$h$ & $0.69\pm0.03$ & $0.71\pm0.034$ \\
$\tau$ & $0.155\pm0.057$ & $0.143\pm0.054$\\
$F_b$& $1.46^{+0.24}_{-0.22}$ & $1.47^{+0.2}_{-0.18}$\\
\hline
\end{tabular}
\end{center}
\caption{\label{tab:results} The best fit results}
\end{table}

Each of our eight chains allowed variation in all nine parameters.
The results are summarized 
in Fig.~\ref{allmarg} and 
Table~\ref{tab:results}. In Fig. \ref{allmarg} we plot the marginalized 1-D 
posterior distribution functions for the nine parameters we allowed to vary 
in our analysis.
The solid lines represent the PDFs for these parameters with 
cosmic strings included; the dashed lines show the results without cosmic strings 
($B=0$). We have shaded the regions excluded at 68\% (light) and 95 \% (dark) 
confidence. The peaks of each of these PDFs and the one-sigma error bars 
are given in Table \ref{tab:results}; for the parameter $\tau$, where the results lack a clear
peak, we have taken the midpoint of the 68 \% confidence region as our 
``peak" value. 


The resulting cosmology with cosmic strings included is very close to the cosmology 
without cosmic strings.  This verifies our hypothesis: a subdominant admixture 
of cosmic strings into the cosmological perturbation spectra gives a minor 
modification to the resulting cosmological parameters as determined in such a model.  This justifies 
the approximation made in \cite{PTWW03,PWW04}, where we fixed all of the cosmological parameters 
except for $n_s$ to their WMAP best-fit values. The only noticeable shift caused by including cosmic 
strings comes in the peak likelihood in the PDF for $\Omega_B h^2$, and even 
this shift is small. The only potentially worrisome aspect of these PDFs is the fact 
that the PDF for $\tau$ appears to be running into the upper value of our prior range.

\begin{figure}
\centering
\includegraphics[width=80mm]{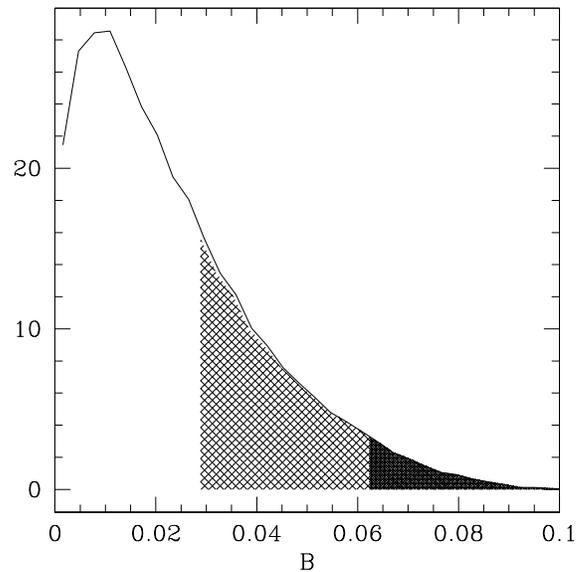}
\caption{\label{bmarg} The PDF for the Cosmic String weighting parameter,
 $B\equiv (G\mu / 2 \times 10^{-6})^2$. }
\end{figure}

\begin{figure}
\centering
\includegraphics[width=80mm]{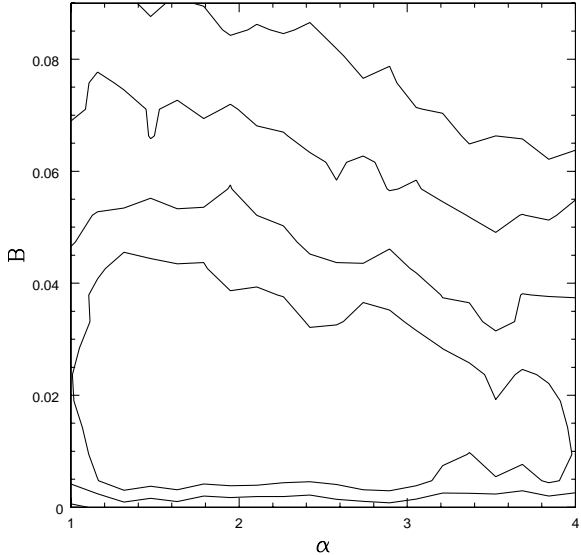}
\caption{\label{bacont} A contour plot of a separate set of MCMC results in the 
two-dimensional space defined by the parameters $B$ and $\alpha$. The contours are somewhat 
crude, but suggest that $\alpha$ is poorly constrained. At this level, there is
no evidence for degeneracy between the parameters. }
\end{figure}

                           
Let us examine the PDF for the cosmic string weighting parameter, $B$, 
given in Fig. \ref{bmarg}.
The light and dark shaded regions again represent
the 68 and 95\% 
confidence intervals. Although the parameter name, $B$, is retained from our previous work \cite{PWW04}, 
the fact that we have changed our string model has changed this parameter's meaning. 
Thus, Fig. \ref{bmarg} 
should not be directly compared with our earlier results. 
Indeed, we have added a new parameter for describing the strings, $\alpha_r$, the string wiggliness. 
In our previous code, we had effectively fixed $\alpha_r = 1$.  Since strings with $\alpha_r > 1$ have a 
larger effective mass per unit length than $\alpha_r =1$ strings, the cosmological
perturbations caused by $\alpha_r>1$ strings are generally larger. Note that
$\alpha_r$ and $B$ were not degenerate parameters -- the effect of a higher $\alpha_r$, 
for a particular spectrum, was quite independent of $B$; see a (rough) contour plot in 
Fig. \ref{bacont}. This is because varying the wiggliness modifies the shape
of the spectra in addition to their amplitude, as we described briefly in \S \ref{stringmodel} \cite{PV99}.

\begin{figure}
\centering
\includegraphics[width=80mm]{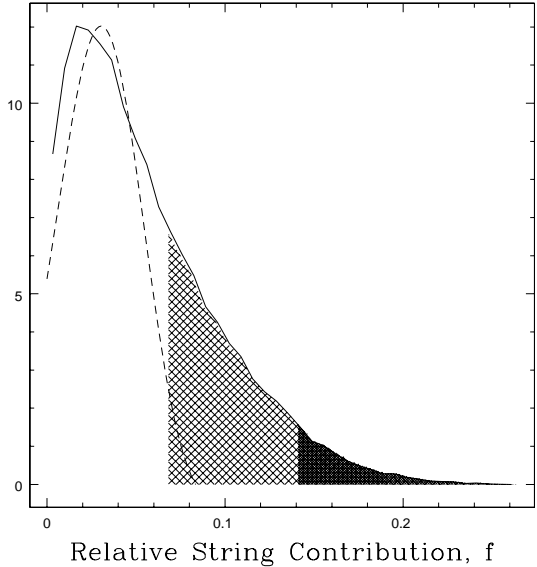}
\caption{\label{fmarg} The PDF for the combination parameter $f$, which quantifies 
the fractional contribution of cosmic strings to the total $C_l^{TT}$ spectrum. 
The solid line shows $f$ from our full analysis, with the 68 \% (light) and
95 \% (dark) confidence regions shaded. The dotted line is the result for $f$ from our previous 
three parameter analysis.
}
\end{figure}

To quantify the total string contribution to the CMB anisotropy for a given set of 
parameters $\vec{s}$ we define the fraction $f$ as
\be
f(\vec{s}) = \frac{I^{cs}}{I^{cs} + I^{ad}} \,
\label{definef}
\ee
where
$$
I^{cs} = \sum_l \frac{(2l+1)}{4\pi} C_l^{\rm cs}(\vec{s}) \ ,
$$
and 
$$
I^{ad} = \sum_l \frac{(2l+1)}{4\pi} C_l^{\rm ad}(\vec{s}) \ ,
$$
where $C_l^{\rm ad,cs}$ are the temperature (TT) correlation spectra.
We then compute the PDF for this new parameter, which we show in Fig. \ref{fmarg} along
with the 68 and 95\% probability regions. We find that about $7$\%($14$\%) of the CMB power 
can be sourced by strings at $68\%$($95\%$) confidence. Note that the dashed line,
corresponding to the three parameter analysis of \cite{PWW04}, shows a 
smaller allowed string contribution. 


Figs. \ref{bmarg} and \ref{fmarg} are the principal results of this paper. Limits on $B$ 
alone give limits on the string tension itself. Using the
results in Fig. \ref{bmarg}, we find a cosmic string weight of $B \lesssim 0.029(0.06)$ allowed
at the $68$ ($95$)\% confidence level. This corresponds to $G\mu \lesssim 3.4 (5.0) \times 10^{-7}$. 
The peak of the PDF for $B$ lies at $B=0.01$, or $G\mu \sim 2 \times 10^{-7}$. These limits are 
relevant to searches for direct detections of cosmic strings, as the magnitudes of 
gravity wave and lensing events caused by cosmic strings depend directly on their tension.


The above bounds on $G\mu$ are model-dependent. In order to match the
total observed CMB power, our string model requires
$G\mu_0 \sim 2\times 10^{-6}$. This value is on the upper end of other estimates in the literature,
which means that our upper bound on $G\mu$ can be treated as a conservative upper bound.
Our $G\mu_0$ is consistent with the COBE normalized values in \cite{allen97}
($G\mu \approx 1.7 \times 10^{-6}$), in \cite{bennett92} 
($G\mu=(1.5 \pm 0.5)\times 10^{-6}$), in \cite{leandros93} ($G\mu=(1.7 \pm 0.7)\times 10^{-6}$), and
in \cite{coulson94} ($G\mu \approx 2\times 10^{-6}$). 
It also agrees with  \cite{ABR97} (for similar model parameters). 
Significantly different estimates of $G\mu_0$ were found
in \cite{carlo99}, where the COBE normalized value of $G\mu$ was 
$\sim 1.0 \times 10^{-6}$ (for their parameter $w^X=1/3$), and similarly in \cite{dani}.
The latest estimates of Landriau and Shellard \cite{LS04} using realistic simulations of 
cosmic strings \cite{LS04} (reliable up to $\ell \sim 20$) give the COBE normalized value of 
$G\mu = (0.74 \pm 0.2) \times 10^{-6}$ for a $\Lambda$CDM cosmology, which is consistent with
results of a similar study in \cite{allen96} where 
the value obtained was $G\mu = (1.05 \pm 0.3) \times 10^{-6}$. Note that most of the
results obtained prior to $1999$ assumed a CDM dominated cosmology -- switching to $\Lambda$CDM 
leads to a $\sim 10$\% increase in COBE normalized value of $G\mu$ \cite{LS04}.

Our bound on $G\mu$ would also be altered if the strings intercommute at a rate
less than unity, as is expected in many string theory models of cosmic strings. The effect
of reduced intercommutation would be to lower the upper limit on $G\mu$.

Our bound on the fraction of CMB power in strings, $f$, depends only on the {\it shapes} of
the string-sourced CMB and LSS spectra. These shapes, as discussed in \S \ref{stringmodel}, 
are largely independent of the details of the string model. The bound on $f$ can be used to
derive an approximate bound on 
$G\mu$, given one's favorite value of $\mu_0$. For example, if one accepts the values 
in \cite{allen96,LS04}, i.e. $G\mu_0 \sim 10^6$, 
one obtains  $G\mu' \sim f^{1/2} G\mu_0 \lesssim 2 (3) \times 10^{-7}$ at $68$ ($95$)\% confidence level.

It is also worth recognizing that $f$ can serve as a
measure of the goodness-of-fit of the paradigmatic inflationary scenario
in comparison with a physically motivated model; isocurvature is another example
of a model used in such a manner (e.g. \cite{isocurvature}).
Our results from cosmic strings, serving from this viewpoint merely 
as a self-consistent foil to the standard model, 
show that as much as 14 \% of the CMB TT-correlation power could be sourced by a radically 
different spectrum without destroying the close agreement of the resulting spectrum with the
anisotropy data. Loosely speaking, we can conclude from this bound that there is a cumulative 
ambiguity in the uniqueness of the adiabatic $C_{\ell}$ spectrum shape, 
as determined from the WMAP data, of around 10\%. As more CMB data become available in the
future, repetition of this analysis might be worthwhile, if only to discover whether 
the intrinsic shape of the adiabatic spectrum is more uniquely picked out by the more-exact
future data sets.

\subsection{The B-polarization spectrum}

\begin{figure}
\centering
\includegraphics[width=80mm]{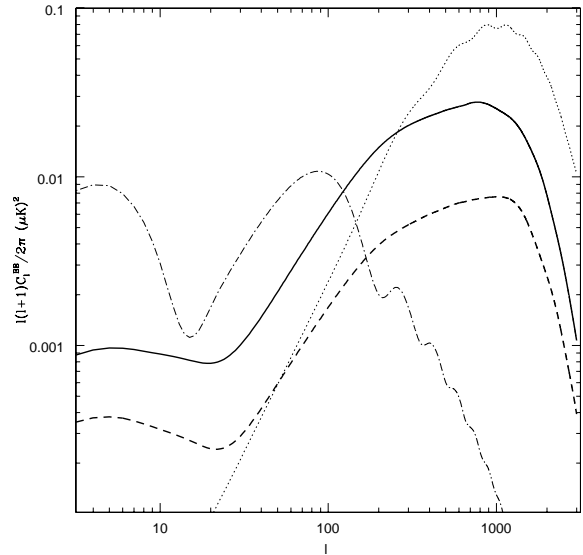}
\caption{\label{clbb} The B-type polarization spectra for $\alpha_r=1.9$ (dashed lined) and
$\alpha_r=1$ (solid line) both corresponding $f=0.1$. The light dotted line is the B-mode 
expected from gravitational lensing of adiabatic E-mode polarization; the light dash-dotted line
is the B-mode expected from gravity wave-sourced polarization, for tensor-to-scalar ratio $r=0.1$.}
\end{figure}

In Fig.~\ref{clbb} we plot the B-mode polarization spectra in the case
of smooth strings ($\alpha_r=1$) and wiggly strings $\alpha_r=1.9$ predicted
by our string model for the case when the total contribution of strings
to the CMB anisotropy is $10$\%. That is, for each of the curves,
the value of $G\mu$ was adjusted separately to correspond to $f=0.1$.
For comparison, we also plot the B-mode spectra from a purely adiabatic cosmology.
The light dotted line represents the B-mode polarization arising from gravitational lensing 
of E-mode polarization. The light dash-dotted line represents the B-mode 
arising from gravitational waves. Any direct detection of cosmic-string generated B-modes 
will rely upon accurate predictions of the spectrum of B-mode polarization 
arising from lensing of E-mode polarization. Fortunately, several 
observational groups believe that it will be possible to ``clean" as much as 90 \%
of the apparent E to B lensed power through accurate E-mode observations.
With gravity-wave generated B-modes peaking at a much lower $\ell$, 
any excess power in observed B-mode spectra at $\ell \sim 1000$ could be 
a telling sign of cosmic string activity. Two planned experiments, QUIET and 
QUaD \cite{polarization}, expect to be able to measure such high-$\ell$ polarization 
with great precision. It is also worth noting that the amplitude of gravity-wave
generated B-mode polarization is intimately tied to the as-yet undetermined 
scalar-to-tensor ratio, $r$. We have used $r=0.1$ in Fig. \ref{clbb}, which is
usually regarded as an optimistically high value. Inflationary models frequently 
produce orders-of-magnitude lower estimates for $r$ (in \cite{hassan}, for instance,
investigations predict $10^{-8} \lesssim r \lesssim 10^{-3}$ for 
KKLMMT-motivated brane inflation). For $r \ll 0.1$, 
cosmic strings could be the dominant source of B-mode polarization for low values of 
$\ell$ ($10 < \ell < 1000$), but with a spectrum that is recognizably distinct from the polarization 
generated by gravity waves. The proposed CLOVER experiment \cite{clover}
and its space-based successors plan to focus their measurements on this region of $\ell$-space.

\section{Summary}
\label{summary}

We found two types of constraints on cosmic string networks
through our analysis of the WMAP and SDSS data. One is on the value of $G\mu$, which
is sensitive to our normalization convention and to the string intercommutation rate, which
we take to be unity in setting our bound. 
The other is on the fraction of total CMB power due to cosmic strings, $f$.
The constraint on $f$ depends chiefly on the general shape of the string induced spectrum.
The shape of the spectrum is quite generic -- a plateau on
large scales due to superposition of the Kaiser-Stebbins 
effect of many strings (from a scaling network), followed by
a broad peak on small scales. The peak is caused by 
the Doppler and Sachs-Wolfe effects produced by velocity and density
perturbations caused by strings during
the epoch of recombination. The rough position of the peak is set by
the size of a typical wake at last scattering, while the wake size is set by the
coherence (or curvature) length $\xi(t)$ at that time. We can call these features 
generic because they are agreed upon by all groups that have studied
string-induced CMB spectra.
The fine details of the shape, e.g. the sharpness of the peak, depend on many factors.
Among the most important of those extra factors is the amount of small-scale structure 
on strings (the wiggliness), which we included as a parameter in our calculation. 

In addition to quantifying the allowed fraction of the cosmic string contribution,
our parameter $f$ can also be interpreted
as a measure of the goodness-of-fit of the fiducial adiabatic CMB spectrum model. The fact that 
the data permit approximately 10\% of the TT-correlation power to arise from a very 
different competitor model gives a useful measure of how uniquely the data
pick out the shape of the adiabatic spectrum. 
 
Other recent constraints on local strings that also used WMAP data include
ref. \cite{Wu}, where the narrowness of the first peak was used to constraint
the size of the incoherent string contribution, and refs. \cite{smoot,wright}, where
the constraint was based on the expected non-Gaussian signatures induced
by strings. Ref.~\cite{Wu} suggests an interesting way to obtain a rough
bound the string contribution to the CMB spectrum at $\ell \sim 220$ from 
constraints on the width of the main peak.
Our method has the advantage of including
the information on all scales (not just the main peak) and the ability
to account for changes in in cosmological parameters and the shape of the string spectra (by
varying the wiggliness). 

Our limit on the string tension, $G\mu < 5 \times 10^{-7}$, does not 
contradict a variety of recent claims of observational evidence for
the existence of cosmic strings. The most prominent have been 
possible examples of cosmic string lensing. In the first case
 \cite{sazhin}, a pair of nearly identical images were found, 
 the best explanation of which appears to be a direct lensing
 by a cosmic string with tension $G \mu \sim 4 \times 10^{-7}$.
Another possible cosmic string lensing observation \cite{Schild}
is the appearance of short time-scale variations in the 
brightness of the well-known gravitational lens system, 
Q$0957+561$, which could be explained by a passing
cosmic string loop with a tension in the range $10^{-8} \leq G\mu \leq 6 \times 10^{-7}$.
The inferred time scale of these variations is so short, $\sim100$ 
days, that very small scale string structure would be required, making the
claim somewhat problematic (for more on these events, see \cite{Kibble04}).
It is also possible that strings are responsible
for the early reionization suggested by WMAP \cite{wmap_kogut}, since cosmic
string wakes can bring about star formation much earlier \cite{rees}
than the standard Press-Schechter scenario \cite{Press}; conservative analysis of
string-mediated early reionization suggests a bound of $G\mu \lesssim 5\times10^{-7}$
\cite{PV04}.
In the arena of gravitational wave observation, current pulsar timing bounds are
still marginally consistent with $G \mu \sim 10^{-7}$ \cite{timing}, while analysis
of gravitational wave bursts from string loops suggests that $G\mu \sim 10^{-7}$
cosmic strings will be readily observable by both LIGO and LISA \cite{DV04}.

It is often thought that a key observational test for cosmic strings would be
whether non-Gaussianity is found in the primordial perturbations seen by such 
experiments as WMAP. This is a natural assumption, since each string acting alone
would perturb the CMB in a highly non-Gaussian manner. However, 
the central limit theorem tells us that the superposition of
perturbations produced by many strings must be Gaussian.
Therefore, one expects to see string-related non-Gaussian 
features only on scales that are sufficiently small not to have been crossed
by more than a few strings during the entire period of time during which 
strings have produced their effects.
It is not difficult to get a conservative estimate of this scale.
The dominant contribution to the anisotropy on small (sub-degree) scales comes 
from the Doppler and Sachs-Wolfe effects produced at the last scattering \cite{leandros95}. 
A natural length scale to start with is the angular size of the horizon at recombination, 
which corresponds to $\ell \sim 220/\sqrt{3}$. 
Numerical simulations \cite{BenBouch,AllenShellard90} show
that the typical distance between strings during matter domination
is $L\sim 0.8t$. It is somewhat smaller in the radiation era, and can be much smaller
if the intercommutation probability was less than unity, as may be 
the case in string theory models. A conservative estimate of the number of
strings per horizon at any time in the matter era is
$\sim (4\pi t^3/3)/L^3 \sim 10$. When projected onto the last scattering surface,
about half of this number of strings would contribute. Hence, the CMB anisotropy in a patch
corresponding to $\ell \sim 220/\sqrt{3}$ would be a superposition of the effects
of about $5$ strings, and to isolate the effect of one string, one would have
to go to scales of order: $\ell \sim 220 \times 5/\sqrt{3} \sim 600$.
In doing this rough estimate we have ignored the density
perturbations created by wakes between the radiation-matter equality and
last scattering, which contribute a non-negligible fraction of the power 
near the main peak \cite{leandros95}. This contribution would tend to make the 
scale at which non-Gaussianity appears even smaller, since the wakes started
to from from the onset of matter domination.
Thus one will likely need a resolution of at least $\ell \sim 1000$
to have any hope of seeing non-Gaussianity from strings.
Furthermore, in the above argument we did not account for the fact that
strings can produce only $10$\% of the total anisotropy. This makes the
detection of their non-Gaussian signatures even more difficult. The possibility
of low string intercommutation rates (high string density), in addition to 
making strings more Gaussian (via the central limit theorem), also strengthens the
bound on their tension, hence further complicating the detection of
their non-Gaussian properties.
The analysis in refs. \cite{smoot,wright} assumed $\ell\sim 200$ 
as the scale for the onset of non-Gaussianity in the CMB caused
by strings. A more realistic scale, as we have argued above, is likely to be an order 
of magnitude smaller, so the detection of string sourced non-Gaussianity appears to be 
beyond the reach of WMAP, and quite possibly even Planck. 
The existing constraints on string-sourced 
CMB non-Gaussianity, such as those obtained in refs. \cite{smoot,wright}, appear to reflect the 
limitation that, given the variance of a CMB map on a certain scale, $\sigma_\ell$, 
one naturally has difficulty resolving any detailed features on those scales
that have amplitudes comparable to $\sigma_\ell$. In light of this, it is not surprising
that the upper bound on $G\mu$ obtained in \cite{smoot,wright} ($G\mu \lesssim 10^{-5}$)
roughly corresponds to the variance of the WMAP CMB map on sub-degree scales.

We find that cosmic strings with tensions of $G\mu = 5 \times 10^{-7}$ are still 
allowed by the data from the WMAP and SDSS experiments, and that 
strings can account for as much as 14\% of the the temperature fluctuations in
a cosmic microwave background radiation dominated by adiabatic fluctuations
without any significant changes in the underlying cosmology. Indeed, 
the excess power at very small angular scales seen in CMB observations 
may already suggest the presence of cosmic string-generated perturbations,
which may dominate on such scales \cite{PTWW05}. Strings with 
allowed tensions are produced in brane inflation models, implying that such
models are still viable and that the strings produced by them may be observable,
giving us hope of an observational window on string theory. One promising
signature of cosmic strings with these tensions in the early universe would be their
creation of observable B-mode polarization in the CMB with spectra
distinct both from those created by E-mode lensing and by gravity waves
from primordial tensor modes. If primordial tensor modes are weak, cosmic 
strings could be the principal new physics seen by B-mode observations;
successful observation of such B-modes would in turn be our first direct
observational probe into the physics of string theory.

\acknowledgments
We thank Ken Olum, Leandros Perivolaropoulos, Paul Shellard, Henry Tye, Tanmay Vachaspati
and Alex Vilenkin for helpful comments and discussions, Licia Verde for assistance 
with using the WMAP likelihood code, Max Tegmark for writing an easy 
to use SDSS likelihood code and making it publicly available
\cite{maxwebsite}, and Marina Romanova and Bruce Wyman
for loaning us computational facilities. This research is partially supported by NSF
Grant No. AST 0307273 (I.W.). M.W. is supported by the NSF Graduate
Fellowship.

\end{document}